\documentstyle[12pt]{article}
\def\edoc{\end{document}}

\def\arcsinh{\mathop{\rm arcsinh}\nolimits}
\def\arccoth{\mathop{\rm arccoth}\nolimits}
\def\csch{\mathop{\rm csch}\nolimits}
\def\grant{No 2 P03D 014 17}
\def\mhalf{^{-1/2}}
\def\nn{{\bf n}}
\def\xx{{\bf x}}

\def\kk{{\bf k}}
\def\hh{{\bf h}}
\def\be{\begin{equation}}
\def\ee{{\end{equation}}}
\def\ygr{\widehat{X}}
\def\ss{p}
\def \Journal#1#2#3#4{{#1} {\bf #2}, #3 (#4)}

\def \NPB{{\em Nucl. Phys.} B}

\def \PRD{{\em Phys. Rev.} D}

\def \ApJ{\em Astrophys. J.}

\def \AA{\em Astronomy Astrophys.}
\def \RMPh{\em Rev. Mod. Phys.}

\def \CQG{\em Class. Quantum Grav.}

\def \JETP{\em Sov. Phys. JETP}
\def \ZETP{\em Zh. Exp. Theor. Fiz.}
\def \JP{\em J. Phys. \rm(USSR)}
\def \AP{\em Adv. Phys.}

\def \MNRAS{\em Mon. Not. R. Astron. Soc.}

\def \PTP{\em Prog. Theoret. Phys.}
\def \PTPS{\em Prog. Theoret. Phys. Suppl. No.}

\begin{document}

\title{Acoustics of early universe.\\
--- Flat versus open universe models}

\author{Zdzis{\l}aw A. Golda and Andrzej Woszczyna\\
Astronomical Observatory, Jagellonian University\\
ul. Orla 171, 30--244 Krak\'ow, Poland}
\date{\today}

\maketitle
\begin{abstract}
A simple perturbation description unique for all signs of curvature,
and based on the gauge-invariant formalisms is proposed to demonstrate
that:
 (1) The density perturbations propagate in the flat
radiation-dominated universe in exactly the same way as electromagnetic
or gravitational waves propagate in the epoch of the matter domination.
 (2) In the open universe, sounds are dispersed by curvature.
The space curvature defines the minimal frequency $\omega_{\rm c}$ below
which the propagation of perturbations is forbidden.

Gaussian acoustic fields are considered and the curvature imprint in the
perturbations spectrum is discussed.
\end{abstract}


\newpage

\section{Introduction}
\label{intro}

Discovering the wave nature of scalar perturbations in the early
universe has a long history. Watchful reader of Harrison's classical
paper~\cite{harrison} can guess the wave equations out of formulae given
there (Section 5.5). Trigonometric or Bessel solutions together with
$\omega\eta$-dependence characteristic for flat perturbed universes
appear in both classical~\cite{sakharov,sakai} and gauge-invariant
theories~\cite{bardeen}. In the case of flat universe the gauge-specific
wave equations are explicitly given by Sachs and Wolfe
(\cite{sachs&wolfe} see theorem pp.~76--77). The comprehensive phonon
description of perturbations in the flat radiation-filled universe,
together with the attempt to quantize them, has been formulated by
Lukash~\cite{lukash} and continued in its quantum aspect by
others~\cite{chibisov&mukhanov}. The wave character is
confirmed~\cite{golda&woszczyna} in the original
Lifshitz-Khalatnikov~\cite{lifshitz&khalatnikov} formalism. Acoustic
motions of the baryon-electron system after recombination have been
noticed by Yamamoto {\it at al.}~\cite{yamamoto&sugiyama&sato}. Some
parallels between scalar perturbation dynamics and gravitational waves
can be found in~\cite{grishchuk,white}.

Controversies, however arise over the gravitational instability
criteria, the gauge problems and the role of the space curvature.

(1) In the $\eta\rightarrow 0$ limit one can formally construct the
growing and decaying solutions. Since these solutions are typically
considered as the large scale approximation ($\omega\rightarrow 0$) the
structure formation is expected in scales greater than the sound
horizon. Consequently the Jeans criterion is understood as the
dispersion relation dividing perturbations into two classes: acoustic
waves and gravitationally bound structures. No dispersion relations like
that can be inferred from the exact solutions \cite{sakai}-\cite{lukash}, 
\cite{golda&woszczyna}.

(2)~As long as the results depend on the coordinate system (the gauge-specific 
solutions \cite{sachs&wolfe,lukash,golda&woszczyna} differ one
from another) their physical meaning is a subject of dispute.
Acoustic field deserves complete gauge-invariant treatment.

(3)~The problem of acoustic field does not seem to be solved properly
in open universes, where most authors traditionally employ the flat
space Fourier analysis, instead of Fourier expansions in the Lobachevski
space.

In attempt to clear those points, we propose a simple perturbation
description, which is unique for all signs of curvature, and based on
the gauge-invariant perturbation formalisms~(Sakai~\cite{sakai},
Bardeen~\cite{bardeen}, Kodama and Sasaki \cite{kodama&sasaki}, Lyth and
Mukherjee~\cite{lyth&mukherjee}, Padmanabhan~\cite{padmanabhan},
Brandenberger, Kahn and Press~\cite{brandenberger}, Ellis, Bruni and
Hwang~\cite{ellis&hwang&bruni}, Olson~\cite{olson}, see also
\cite{woszczyna&kulak}).

Section \ref{scalar} contains a brief recipe of how to reduce equations
obtained in these theories to a single, second order partial
differential equation~(\ref{partial0}). Differences between the
formalisms occur to be of no importance here, and we obtain exactly the
same propagation equation for all of them. We show how to transform this
equation to the wave equation in its normal form.

We obtain a general, ``profile-independent'' solution for the flat
universe (Section~\ref{flat}), without appealing to the Fourier
transform. We demonstrate that the gauge-invariant density perturbation
propagate in radiation-domi\-na\-ted universe in the same way as
electromagnetic or gravitational waves propagate in the epoch of the
matter domination. Eventually, we expand perturbations into
planar waves, in order to discuss some basic features of the spectrum
and the spectrum transfer function.

In section \ref{fourier} we describe the sound propagation in open
universes. We analyse the dispersive role of the curvature. The space
curvature prevents perturbations of frequencies smaller than some
critical $\omega_{\rm c}$ from propagating in space, and systematically
reduces the group velocity for others, when $\omega$ goes down
to~$\omega_{\rm c}$.

Section \ref{acoustic} is devoted to Gaussian acoustic fields. We derive
the spectrum transfer function in the form suitable to estimate the role
of the space curvature in the microwave background.

\section{Scalar perturbations in the early universe}
\label{scalar}

In the universe filled with highly relativistic matter the
energy momentum tensor is trace-free. The dynamics of the scale
factor $a(\eta)$ expressed as a function of the conformal time
$\eta$ is governed by
	\begin{equation}
T^{\mu}_{\phantom{\mu}\mu} =-\frac{6}{a^3(\eta)}
\left(a''(\eta)+K\,a(\eta)\right)=0 \label{friedman}
\end{equation} and yields \begin{equation} a(\eta)=
\sqrt{\frac{{\cal M}}{3}}
\frac{\sin\left(\sqrt{K}\eta\right)} {\sqrt{K}}.
	\label{scaleterm}
	\end{equation} 
We treat the curvature index $K$ as continuous quantity and keep
$K$ explicitly in both the equations and solution, as far as
possible. Traditional formulae can be recovered by setting
$K=\pm1$ or by the limit procedure $K\to 0$. Normalization
$\sqrt{{\cal M}\slash3}$ recalls the constant of motion
${\cal M} = \rho(\eta)a^4(\eta)$. 

The perturbation equation expressed in the orthogonal gauge\footnote{or
the gauge-invariant differential measures of inhomogeneity
\cite{ellis&hwang&bruni}},~\cite{olson,woszczyna&kulak} and
parameterized by the conformal time\footnote{In the orthogonal gauge the
conformal time is defined as the integral $\eta = \int\!\frac{1}{a(t)}
{\rm d}t$, where time $t$ means the orthogonal time --- the time
parameter constant on orthogonal hypersurfaces.} $\eta$ takes the
canonical form (free of first derivatives) \begin{equation}
\frac{\partial^2}{\partial\eta^2}X(\eta ,\xx)-\frac{2{\cal
M}}{3{{a^2(\eta)}}} X(\eta ,\xx) -\frac{1}{3} \,{}^{(3)}\!\Delta X(\eta
,\xx)=0 \label{partial0} \end{equation} where ${}^{(3)}\!{\Delta}$
denotes the Laplace-Beltrami operator acting on orthogonal
hypersurfaces.

Equation (\ref{partial0}) can be easily derived from the
Raychaudhuri and the continuity equations (see the procedure
in~\cite{lyth&mukherjee} or~\cite{padmanabhan}). It also can be
recovered from the Sakai equation~\cite{sakai} (formula~(5.1),
$K\to X$), the equation for density perturbations in orthogonal
gauge (Bardeen's~\cite{bardeen} formula~(4.9), $\rho_{\rm
m}\to X$, Kodama and Sasaki~\cite{kodama&sasaki} chap. IV,
formula~(1.5), ${\mit\Delta}\to X$, Lyth and
Mukherjee~\cite{lyth&mukherjee} formulae (16--17), $\delta
\to X$, Padmanabhan~\cite{padmanabhan} Eq.~(4.88), $\delta\to X$), the
equation for gauge invariant metric potentials (Brandenberger, Kahn and
Press~\cite{brandenberger} formula~(3.35), ${\mit\Phi}_H\slash\rho
{a^2} \to X$), the equation for gauge invariant density gradients
(Ellis, Bruni and Hwang~\cite{ellis&hwang&bruni} formula~(38), ${\cal
D}\to X$) or Laplacians (Olson~\cite{olson} formulae~(8--9), as well as
its extension to open universes~\cite{woszczyna&kulak} formula~(22))
after transforming these equations to conformal time (if parameterized
differently) and employing the Helmholtz equation to restore partial
form of the perturbation equation. Suitable changes of the variable
names as indicated above ($\mbox{\it original\/}\to X$) are necessary.

We introduce a new perturbation variable $\ygr$

	\begin{equation}
\ygr(\eta,\xx)=\frac{1}{a(\eta)} \frac{\partial}{\partial\eta}(a(\eta)
X(\eta,\xx)).
	\label{var_Y}
	\end{equation}
While $X(\eta,\xx)$ satisfies (\ref{partial0}) and
$a(\eta)$ is given by (\ref{scaleterm}) the perturbation
variable~$\ygr$ obey the wave equation in its normal form
	\begin{equation}
\frac{\partial^2}{\partial\eta^2}\ygr(\eta,\xx) -\frac{1}{3}\,
{}^{(3)}\!\Delta \ygr(\eta ,\xx)=0.
	\label{normal}
	\end{equation}

The time derivatives of the gauge-invariant inhomogeneity measures also
form gauge-invariant variables. In this sense variable $\ygr$ is equally
good as $X$. However, time derivatives may be difficult to observe at
the last scattering surface, and hardly represent physically
meaningful aspects of the cosmic structure. The equation
(\ref{normal}) plays only an auxiliary role, nevertheless is very
useful. Formally, it describes the wave (massless field) propagating in
the static space-time of constant space-curvature. The specific case of
positive curvature (Einstein static universe) has been
considered~\cite{birrell&davies} in the context of quantum field theory
on curved background. In our case spaces of zero or negative curvature
are of particular importance. We will discuss both cases individually.

\section{Sound waves on the flat background}
\label{flat}

When the space curvature vanishes the equation (\ref{partial0}) reads as
        \begin{equation}
\frac{\partial^2}{\partial\eta^2} X(\eta,\xx) 
-\frac{2}{{\eta }^2} X(\eta,\xx) -
\frac{1}{3}\, {}^{(3)}\!\Delta X(\eta,\xx)=0
	\label{partial1}
	\end{equation}
and is essentially the same as the propagation equation for
gravitational~\cite{grishchuk,white} or
electromagnetic\footnote{See formula (5.2.6) in \cite{weinberg}
after substituting $g=\eta^{6}$.} waves in the dust-filled
universe
	\begin{equation}
\frac{\partial^2}{\partial\eta^2} X(\eta,\xx) -\frac{2}{{\eta }^2}
X(\eta,\xx) -{}^{(3)}\!\Delta X(\eta,\xx)=0.
	\label{gw1}
	\end{equation}
The only differences are that gravitational and electromagnetic waves
are expressed by the tensor $h_{\mu\nu}$ and vector $A_{\mu}$
respectively, and they propagate with the speed of light ($c=1$), while
the solutions to equation~(\ref{partial1}), represent scalar waves
travelling with the velocity $v=1/\sqrt3$.

Now, the Laplacian ${}^{(3)}\!\Delta$ operates in Euclidean space.
Equation (\ref{normal}) when expressed in Cartesian coordinates
$\{\xx\}$ is solved by an arbitrary function $\ygr=\ygr(\nn{\cdot}\xx- v\eta)$
\cite{witham} with $v=1\slash\sqrt3$ and $\nn{\cdot}\nn=1$. However, to keep
the linear approximation valid we require that $\ygr(\nn{\cdot}\xx- v\eta)$ to be
limited throughout the space\footnote{$\forall\eta\,\exists{\epsilon}\in
\mbox{I{\kern-.15em}R}\colon~\forall \xx\phantom{x} -{\epsilon}<\ygr(\nn{\cdot}\xx-
v\eta)<{\epsilon}$.}.

Knowing the general solutions of the equation (\ref{normal}) we can
return to observables $X$. We look for the general form of $X(\eta,\xx)$
among the solutions of the equation (\ref{var_Y})
	\begin{equation}
X(\eta,\xx)
=\frac{1}{\eta}\left(\int_0^{\eta}\!\eta'\, \ygr(\nn{\cdot}\xx- v\eta'){\rm d}
\eta' +F(\xx)\right)\!.
	\label{sol1}
	\end{equation}
On the strength of (\ref{partial1}) $F(\xx)$ is harmonic
${}^{(3)}\!\Delta F(\xx)=0$ and must be constant if limited
throughout the space of constant curvature~\cite{stewart}. With
no loss of generality\footnote{The freedom to choose this
constant is not different from ambiguity in the indefinite
integral in (\ref{sol1}). Appropriate integration constants are
traditionally tuned to give the perturbation spatial average
equal to zero.}, we put $ F(\xx)=0$.
Eventually, the general, spatially limited solution to the equation
(\ref{partial1}) is expressed by the integral
	\begin{equation}
X(\eta,\xx)
=\frac{1}{\eta}\int_0^{\eta}\!\eta'\, \ygr(\nn{\cdot}\xx -v\eta'){\rm d}
\eta'
        \label{sol2}
	\end{equation}
of an arbitrary, but also spatially limited function $\ygr(\nn{\cdot}\xx- v\eta)$.
The solution describes a wave having the time-dependent profile
and travelling with the constant velocity
$v=1\slash\sqrt{3}$.

This can be easily confirmed by the Fourier expansion analysis. Indeed,
for any real function $\ygr(\nn{\cdot}\xx- v\eta)$ expressed as
	\begin{equation}
\ygr(\eta,\xx)=\int\! ({\tt A}_k {\tt u}_k (\eta,\xx)+ {\tt A}^*_k
{\tt u}^*_k(\eta,\xx)) {\rm d} \kk 
	\label{FY}
	\end{equation}
with
	\begin{equation}
{\tt u}_k(\eta,\xx)=\frac{1}{\sqrt{2\omega}} {\rm e}^{{
i}(\kk{\cdot}\xx-\omega \eta)}
	\label{FYT}
	\end{equation}
corresponds $X(\eta,\xx)$ 
	\begin{equation}
X(\eta,\xx)=\int\!({\cal A} _k\,u_k(\eta,\xx)
+{\cal A}^*_k\, u^*_k(\eta,\xx)){\rm d} \kk
	\label{FXA}
	\end{equation}
expanded into modes $u_k(\eta,\xx)$
	\begin{equation}
u_k(\eta,\xx)
=\mu_{\omega}(\eta){\rm e}^{{i}\kk{\cdot}\xx}
=\frac{1}{\sqrt{2\omega}} \left(1+\frac{1}{{i}
\omega\eta}\right) {\rm e}^{{i}(\kk{\cdot}\xx-\omega\eta)}
	\label{uk}
	\end{equation}
The frequency $\omega$ obeys the dispersion relation
$\omega^2={k^2}\slash3$, the Fourier coefficient ${\cal
A}_k=-\frac{1}{{i}\omega} {\tt A}_k$ is an arbitrary complex function of
the wave number $k$, and $u_k$ is obtained from (\ref{sol2}) after
substituting $\ygr={\tt u}_k$. Modes $u_k$ like ${\tt u}_k$ form an
orthonormal base in the function space with the Klein-Gordon scalar
product~\cite{birrell&davies}. Both ${\tt u}_k$ and $u_k$ form
travelling waves, but only ${\tt u}_k$ have their absolute value
constant in time. Therefore, the generic perturbation $X(\eta,\xx)$ is
composed of plane waves $u_k$ of decaying amplitude, which perfectly
agrees with Sachs and Wolfe results (\cite{sachs&wolfe} pp.~76--77)
obtained in alternative perturbation approach. Waves move with constant
velocity $v=1/\sqrt{3}$ independently of their length-scale. Short-scale and
long-scale perturbations do not form different classes of solutions.

In the theory appealing to stochastic processes the initial perturbation
is given at random at the end of the quantum epoch $\eta_i>0$, and
develops gravitationally according to (\ref{partial1}) in the interval
$\eta>\eta_i$. Therefore, solution's singularity at $\eta=0$ is purely
mathematical fact with no physical consequences.

\section{Sound waves in the curved space}
\label{fourier}

While decomposing perturbations into Fourier series in flat or opened
universes we should respect some specific effects caused by the
curvature. This particularly refers to open universes (the Lobachevski
space), where the orthogonal expansions exist only for a class of
perturbations with sufficiently short-scale
autocorrelation~\cite{caldwell&stebbins}. To expand the others one needs
supplementary series (supercurvature
modes~\cite{yaglom,lyth&woszczyna,langlois,kaiser}) of
non-orthogonal\footnote{In the sense of the Klein-Gordon 
scalar product.} solutions to the Helmholtz equation, which are numbered
by imaginary wave numbers $k\in [-i,i]$.

Let us adopt spherical coordinates $\{r,\theta,\phi\}$ as more
appropriate for curved maximally symmetric spaces. In the same manner
like in the scalar field theories \cite{tanaka} the density perturbation
expands as
        \begin{equation}
X(\eta,r,\theta,\phi)=\sum\limits_{lm}
\int\!(A_{klm}\,u_{klm}(\eta,r,\theta,\phi) 
+A^*_{klm}\,u^*_{klm}(\eta,r,\theta,\phi)){\rm d} k
	\end{equation}
where modes
$u_{klm}(\eta,r,\theta,\phi)=\mu_{\omega,K}(\eta)Y_{klm}(r,\theta,\phi)$ are
expressed by hyperspherical 
harmonics $Y_{klm}(r,\theta,\phi)$ and time-dependent amplitude
$\mu_{\omega,K}(\eta)$ fulfilling the time-equation (obtained by
separation from~(\ref{partial0})):
	\begin{equation}
\frac{{\rm d}^2}{{\rm d}\eta^2}\mu_{\omega,K}(\eta)+ \left(\frac{k^2-K}{3}
-\frac{2K}{\sin^2(\sqrt{K}\eta)}\right)\mu_{\omega,K}(\eta)=0.
	\label{time}
	\end{equation}
We find solutions to (\ref{time}) in the exact form as
        \begin{equation}
\mu_{\omega,K}(\eta)=\frac{1}{\sqrt{2}}\sqrt{\frac{\omega}{\omega^2-K}}
\left(1+\sqrt{K}\,\frac{\cot(\sqrt{K}\eta)}
{{i}\omega}\right){\rm e}^{-{i}\omega \eta}. 
	\label{zet}
	\end{equation}
and their complex conjugates. Solutions $\mu_{\omega,K}(\eta)$ approach
$\mu_{\omega}(\eta)$ (eq. \ref{uk}) in the $K\rightarrow 0$ limit. The
frequency $\omega$ and the wave number are related to each other by the
dispersion relation
         \begin{equation}
\omega(k)=\frac{\sqrt{k^2-K}}{\sqrt3}. 
	\label{dysp}
	\end{equation} 
which can be obtained by simple substitution of (\ref{zet}) into (\ref{time}) 
and perfectly agrees with the dispersion relation obtained for
the variable $Y$ on the strength of equation~(\ref{normal}) 
(compare~\cite{birrell&davies} chapter 5.2).

Functions $Y_{klm}(r,\theta,\phi)$ solve
the Helmholtz equation~\cite{birrell&davies}
	\begin{equation}
{^{(3)}}\!\Delta Y_{klm}(r,\theta,\phi)=-(k^2-K)Y_{klm}(r,\vartheta,\phi)
	\label{partial2}
	\end{equation}
and can be split into the radial part $\Pi_{kl}$ and the two
dimensional spherical functions $Y_{lm}(\vartheta, \varphi)$
		\begin{equation}
Y_{klm}(\chi, \vartheta, \varphi)
= \Pi_{kl}(\chi) Y_{lm}(\vartheta, \varphi)
	\label{separation}
	\end{equation}
Solutions to radial equation 
	\begin{equation}
{\partial^2 \over \partial {\chi}^2} \Pi_{kl}(\chi)
+2\coth \chi {\partial \over \partial {\chi}} \Pi_{kl}(\chi)
-\left(\lambda + {l(l+1) \over {\sinh ^2 \chi}} \right)\Pi_{kl}(\chi)=0
	\label{eq:radial}
	\end{equation}
 are given by
\begin{eqnarray}
\Pi_{kl}&=& N_{kl} \widetilde \Pi_{kl}
\nonumber\\
N_{kl} &=& \sqrt\frac2\pi k^2 \left[ \prod_{n=0}^l (n^2+k^2)
\right]\mhalf
\nonumber\\
\widetilde \Pi_{kl} &=& (k^2\sinh \chi)^l\left(\frac{-1}{k\sinh\chi}
\frac{{\rm d}}{{\rm d}(k\chi)}\right)^{l+1} \cos(k\chi)
\nonumber
\end{eqnarray}
 The lowest multipole solutions
	\begin{eqnarray}
\widetilde \Pi_{k 0}&=& \frac{1 }{\sinh \chi} \left(\frac{\sin {k \chi}}{k}\right)\nonumber \\
\widetilde \Pi_{k 1}&=&\frac{1 }{\sinh \chi} \left (-\cos {k \chi}
+\coth\chi {\sin {k \chi} \over {k}} \right) \nonumber \\
\widetilde \Pi_{k 2}&=&\frac{1}{\sinh \chi} \left (
-3\coth \chi \cos {k \chi }+(3\coth ^2 \chi -k ^2 -1)
{\sin {k \chi} \over {k}} \right)\nonumber
	\end{eqnarray}
are enough to demonstrate properties of both two series of
hyperspherical harmonics. For real wave numbers ($k^2>0$) the $\Pi_{k
1}$ (consequently also $Y_{klm}(r,\theta,\phi)$) functions oscillate in
space. They form an orthonormal basis in the sense of the scalar product
$(f_1|f_2) =\int\!{f_1 f_2^{*}}\sqrt{g}\, {\rm d}^3x.$ As proved by
Gelfand and Naimark they are complete to expand square integrable
functions in the Lobachevski space
\cite{gelfand&naimark,bander&itzykson}. For imaginary wave numbers
contained in the interval $-1\le k^2< 0$, the $\Pi_{k 1}$ (and
$Y_{klm}(r,\theta,\phi)$) functions build supplementary series. These
functions are regular, limited but strictly positive throughout space,
so they are not orthogonal\footnote{Modes with $k=\pm i$ are constant
throughout space. One can subtract them by suitable changes in the
background metric. Other modes with $k\in (-i,i)$, although positive
everywhere, decrease with distance strong enough to keep zero mean
value. For instance, for the spherically symetric perturbation
with the density excess given by 
$\rho(\chi) =
\Pi_{\frac{i}{2} 0}(\chi) = 2\csch(\chi)\sinh\left(\frac{\chi}{2}\right)$
the mass-to-volume ratio 
$\overline{\rho}(r) =\frac{4\pi\int_0^r 
{\rho}(\chi)\sinh^2(\chi){\rm d}\chi}{4\pi\int_0^r 
\sinh^2(\chi){\rm d}\chi}$
tends to zero with volume tending to infinity:
$\overline{\rho}(r)\longrightarrow
\frac{\rho(r)\sinh^2(r)}{\sinh^2(r)}\longrightarrow 
2\csch(r)\sinh\left(\frac{r}{2}\right)\longrightarrow 0.$
No redefinition of the background can absorb perturbations like that.}. The
supplementary series is redundant for expansion of square integrable
functions. Nevertheless, this series is necessary to expand weakly
homogeneous stochastic processes in the Lobachevski
space~\cite{yaglom,lyth&woszczyna}.

In this way in the open universe one obtains two types of hyperspherical
harmonics $Y_{klm}(r,\theta,\phi)$ and consequently two types of modes
$u_{klm}(\eta,r,\theta,\phi)$. Modes $u_{klm}(\eta,r,\theta,\phi)$ with
real $k$ are orthogonal by means of Klein-Gordon scalar
product~\cite{tanaka} and expand waves of square integrable profile.
Modes $u_{klm}(\eta,r,\theta,\phi)$ with $-1\le k^2< 0$ form ``waves of
infinite length-scale''. Both types of modes may contribute to the
spectrum of randomly (or quantum) originated
inhomogeneities~\cite{langlois,kaiser}.

The density perturbations propagate in the open universe in different
manner than the scalar fields or gravitational waves do. Acoustic waves
of different length-scales propagate with different velocities. Indeed
from relation~(\ref{dysp}) we can infer both the phase and group
velocity of sound in the form
	\begin{equation}
v_{\rm f}(k)= \frac{\omega(k)}{k} = \frac{\sqrt{1+k^2}}{\sqrt3 k}
	\end{equation}
and
	\begin{equation}
v_{\rm g}(k) = \frac{\partial}{\partial k}\omega(k) =
\frac{k}{\sqrt3 \sqrt{1+k^2}}.
	\end{equation}
The group velocity decreases with the wave number $k$, to vanish
completely at the $k\to0$ limit. The condition $k=0$ determines the
critical frequency $\omega(0)=1\slash \sqrt3$, below which the wave
propagation is forbidden. Therefore, the acoustic travelling waves are
composed of the principal series modes. The supplementary series build
`global' standing waves of supercurvature scale.

\section{Gaussian acoustic field}
\label{acoustic}

The generic acoustic field described by~(\ref{normal}) is composed of
waves travelling in different directions. Clearly, this property also
refers to solutions of the equation~(\ref{partial0}). The mechanism,
which creates initial small perturbations is expected to be of
probabilistic nature (thermodynamic or quantum fluctuations), therefore
the evolution of linear structure is usually expressed in the language
of stochastic processes. The homogeneity of this stochastic process
reflects the universe homogeneity. Weakly homogeneous processes have
their Fourier expansions
	\begin{equation}
\ygr(\eta,\xx)=\int({\tt A}_k {\tt u}_k(\eta,\xx)
+{\tt A}^*_k {\tt u}^*_k(\eta,\xx))\,{\rm d}\kk
\label{cs1}
	\end{equation}
and consequently
	\begin{equation}
X(\eta,\xx)=\int({\cal A}_k u_k(\eta,\xx)
+{\cal A}^*_k u^*_k(\eta,\xx))\,{\rm d}\kk \label{cs2}
	\end{equation}
where the coefficient ${\cal A}_k$ are random variables of $k$ and the
integral has a stochastic sense~\cite{sobczyk}. In a {\it generic Gaussian
field\/}\footnote{Allen and collaborators~\cite{allen&flanagan&papa}
distinguish between {\it stationary\/} random fields fulfilling
(\ref{e1}-\ref{e2}) and the {\it squeezed\/} random fields, which violates
(\ref{e2}). Stationary processes have their precise meaning in
the framework of the stochastic theory not equivalent to
(\ref{e1}-\ref{e2}). We prefer to name these random fields {\it generic\/}
to stress analogy to generic classical fields.} the expectation values
for ${\cal A}_k$ fulfill
\begin{eqnarray}
E[{\cal A}_k{\cal A}_{k'}^*]&=&{\cal P}_k\delta_{kk'},
\label{e1}\\
E[{\cal A}_k{\cal A}_{k'}]&=&0.
\label{e2}
\end{eqnarray}
${\cal P}_k$~is defined as the field spectrum. The first relation
guarantees the statistical independence of waves of different wave
numbers, the second --- says that no particular phase is preferred.
Waves moving in different directions are statistically independent.

The temperature fluctuations at the last scattering surface
draw our attention to spatial correlations of $\delta\rho/\rho$
measured at the instant $\eta=\eta_{\rm r}$. In the flat universe
the two-point spatial autocorrelation
$R(h)$~\cite{caldwell&stebbins} of the field~$X$ given by
(\ref{zet}) can be expressed as:
	\begin{eqnarray}
R(h)	&=&	\frac{1}{4\pi}\int E[{X}(\xx,\eta){X}
                [\xx+\hh,\eta)] \delta(\hh{\cdot}\hh-1)\,{\rm d}\hh\\
        &=&     \frac{1}{4\pi}\int 2u_ku_k^*{\cal P}_k
                \exp(i\kk{\cdot}\hh)\delta(\hh{\cdot}\hh-1)\,
                {\rm d}\kk\,{\rm d}\hh\\
        &=&     \int_0^\infty 4\pi k^2 j_0(hk)2\mu_{\omega}\mu_{\omega}^*
		{\cal P}_k\, {\rm d}k\\
	&=&	\int_0^\infty 4\pi k^2
\frac{\sin(hk)}{hk}\ss_k(\eta)\,{\rm d} k
	\end{eqnarray}
where $j_0$ is the spherical Bessel function, and $\ss_k$ stands for the
space spectrum of the density perturbation at a given
moment $\eta$. Following Peebles \cite{peebles} we define the transfer function
$T_\omega(\eta)=2\mu_{\omega}\mu_{\omega}^*$, which converts the
time-invariant field spectrum ${\cal P}_k$ into the space spectrum
$\ss_k(\eta)$.
	\begin{equation}
\ss_k(\eta)=T_\omega(\eta){\cal P}_k
=2\mu_{\omega}\mu_{\omega}^*{\cal P}_k.
	\label{t1}	
	\end{equation}
The formula for the space spectrum $\ss_k(\eta)$ splits into
two factors: 1)~$T(\eta)$ --- describing the role of
gravity, and 2)~${\cal P}_k$ coming from other interactions and
rendering their probabilistic nature prior or during radiational
era. We do not discus any specific form of ${\cal A}_k$ or
${\cal P}_k$ in
this paper. We assume, however, that ${\cal A}_k$ enables one to
construct small perturbations, and in particular does not cause
divergences in Fourier integrals. Employing (\ref{uk}) one easily founds
	\begin{equation}
T_\omega(\eta)=\frac{1}{\omega}	\left(1+\frac{1}{(\omega\eta)^2}\right)\!.
	\label{t2}
	\end{equation}
As seen from (\ref{t2}) the evolution of each mode depends on
the product $(\omega\eta)^2$. The contribution from
modes much larger than the horizon scale $\omega\eta\ll 1$
strongly decreases with time, while the perturbation well inside the
horizon $\omega\eta\gg1$ keep constant amplitude. This property confirms
the stability of Robertson-Walker symmetry against the generic (both
classical and stochastic) large-scale density perturbations.

In the same manner one can express random acoustic
fields in the open universe
	\begin{equation}
X(\eta,r,\theta,\phi)=\sum\limits_{lm}
\int\!(A_{klm}\,u_{klm}(\eta,r,\theta,\phi) 
+A^*_{klm}\,u^*_{klm}(\eta,r,\theta,\phi)){\rm d} k.
	\label{14}
	\end{equation}
According to Yaglom's theorem, for weakly homogeneous processes
integration runs over both series principal and supplementary. The
autocorrelation function now reads \cite{lyth&woszczyna}:
	\begin{equation}
R(h)=\int\limits_{R_{+}\cup[0,i]}4\pi
k^2\frac{\sin(hk)}{k\sin(h)}{p}_k(\eta) {\rm d}k
	\end{equation}
and the transfer function determined from formula (\ref{zet})
and (\ref{t1}) takes the form
	\begin{equation}
T_\omega(\eta)=
\frac{1}{\omega}\left(1+\frac{1}{\omega^2-K}
\frac{K}{\sinh^2(\sqrt{K}\eta)}\right)
        \label{to}
	\end{equation}
We rewrite $T_\omega(\eta)$ as a function of the energy density
        \begin{equation}
T_\omega(\rho)=
\frac{1}{\omega}\left(1+\frac{\sqrt{{\cal M}\rho}}{3(\omega^2-K)}\right)
        \end{equation}
to demonstrate that the curvature modifies substantially the space
spectrum $\ss_k$ only in the low frequency limit (supercurvature
modes). Therefore, one may expect to find the curvature signature mostly
in low multipoles (dipole, quadruple)~\cite{langlois}. To extract this
geometrical effect the knowledge of the field spectrum ${\cal P}_k$ is
indispensable.

Expansions (\ref{FXA}) are typically employed in the gravitational
waves theory \cite{white}, and in the scalar field theory \cite{tanaka},
while the density perturbations theories traditionally solve, basically
the same, propagation equation in terms of the Bessel $J_{3/2}$ and
Neumann $N_{3/2}$ functions. The $J_{3/2}$ and $N_{3/2}$ are identified with
``growing'' and ``decaying'' modes respectively, according to their limit
behaviour at $\eta=0$. Since the transition from $\{u,u^{*}\}$ basis to
$\{J_{3/2},N_{3/2}\}$ is the unitary transformation, both
representations are equivalent. When the ``decaying'' mode is rejected
\cite{piki} (a ``standard practice'' in cosmology --- see comments in
\cite{challinor}) the unitarity is broken down and solution space is
truncated to the space of standing waves. Then the acoustic field is in
a highly ``squeezed state'' and consequently the characteristic peaks in the
transfer function appears \cite{piki}.

\section{Remarks on scales and observables}

There is a substantial difference between the dispersion on
curvature, we described above, and ``the curvature imprint'' in
CMBR spectrum, anticipated by the acoustic peaks hypothesis.
The last hypothesis claims that the early perturbed universe was
overdominated by stationary waves \cite{piki}. One
may justify this assumption by appealing to squeezing phenomena
at the transition from deSitter phase to the radiation dominated
epoch. (In a transitions like that the large-scale stationary
gravitational waves are generated (see \cite{abbott&harari}). The
same refers to massless scalar field.) The dominance of
standing waves with specifically correlated phases should
exhibit a series of peaks in the CMBR spectrum. Positions of
these peaks are sensitive to details of the universe dynamics
($\Omega,\Lambda$, etc.). 

On the contrary, the dispersion effect described in the section
\ref{fourier} has strictly geometrical character. It comes directly from
the wave equation in the radiational epoch and does not depend
on universe's past (evolution prior to radiational era is
irrelevant here). The radiation-filled universe become ``opaque''
to sound waves greater than the curvature radius, whatever is
the sound origin. In particular, no additional mechanism
preferring standing waves at the beginning of the radiational era
is needed. On that account the dispersion might form a reliable
curvature tracer, provided it is observable at all, i.e.\\  
1) the space scales of supercurvature perturbations must be
``small enough'' to fit well in the observable part of our
universe, and \\
2) observational data should be complete enough to distinguish
between standing and travelling waves at the last scattering. 

Answer to the first question is relatively easy and it was
already formulated in the literature in terms of multipole
decomposition \cite{langlois}. We repeat the same result below by
use of simple geometrical consideration. Let us assume we live
in the open ($\Omega=0.2$) universe which is presently dominated
by matter ($p=0$) . We see the last scattering surface at some
$\eta_r$ with the redshift $z_r=1000$.
For sake of simplicity (following \cite{abbott&harari}) we assume an
instant transition from radiational epoch ($p=\rho/3$) to
the galactic era ($p=0$), which occur just at the last
scattering moment $\eta = \eta_r$. In such a universe model the scale
factor evolves as 
	\[
a(\eta) = \sqrt{\frac{{\cal M}}{3}}\,\frac{\sinh^2\left(\frac{\eta +
\eta_r}{2} \right)}{\sinh(\eta_r)}
	\]
and the radius of the visible universe $\chi_r$ can be easily
expressed as a function of redshift $z_r$ and the cosmological
parameter $\Omega$ 
	\[
\chi(\Omega, z_r) =
2 \arccoth \left( \frac{1}{\sqrt{1-\Omega}}\right) -2\arcsinh
\left(\sqrt{\frac{1 - \Omega}{\Omega(1 + z_r)}}	\right)
        \]
Setting typical values $\Omega=0.2$ and $z_r=1000$ one obtains
$\chi_r=2.76$. The equator plane ``draws'' on the last scattering
surface a circle of the perimeter $l=2\pi \sinh(\chi_r)=49.5$,
consequently, the curvature radius (in this units equal to one)
takes $\alpha=360{\slash}l =7.28$ degrees on the sky. 
Let us consider now a spherically symmetric density perturbation
(on the last scattering surface) described by the $k=0$
hyperspherical function $Y_ {000} =\chi
\csch(\chi)$. The $Y_ {0lm}$ functions, with arbitrary $l$ and
$m$, form a boundary between 
the principal and supplementary series, and can be understood
as the ``shortest'' supercurvature modes. Since these functions
are positive everywhere, we express the perturbation
length-scale as the half-magnitude width 
$l_{1\slash2}$. For spherically
symmetric mode $Y_{000}$ it is roughly $l_{1\slash2}=2.2$. This
corresponds to $32^\circ$ on
the sky. Equator intersects about 10 patches of that size. The
supercurvature perturbations contribute mainly to the lowest
multipoles in the CMBR spectrum (roughly $l<5$), but the visible
part of the universe is large enough to produce the curvature
effects. Another problem is how to determine from the CMBR data, which
waves are stationary and which of them are travelling ones. The
oscillation time scale for waves close to the critical frequency
$\omega_{\rm c}=\sqrt{1\slash3}$ is of the order of the universe
age, thus no direct observation of their temporal behaviour can
be done. The same refers to gravitational squeezed waves 
(see \cite{allen&flanagan&papa}).
On the other hand, the information we need is ``hidden'' on the
last scattering surface, and the ``only task'' is to read it
properly. For standing perturbations the density and the
velocity fields are strongly correlated 
\cite{ellis&hellaby&matravers}. Density extrema
coincide with the expansion extrema throughout the entire
space\footnote{Considerably more complex coincidences may be expected when
multifluid models are taken into account. These cases need
independent investigations.}. 
This means that even in random acoustic 
fields the density and the velocity perturbations loose their
statistical independence in scales comparable with the space
curvature and consequently would gradually correlate on the sky
when the angle scales outgrow ten degrees. The key to solve this
problem is to find a second independent observable, which would
help us to separate the potential and Doppler contributions to
the CMBR temperature fluctuations. We can hardly propose a
definite candidate at the present stage of observations, but
both the polarization measurements and the large scale flows
analysis seems to be steps in the right direction.

\section{Summary}

The gauge-invariant analysis confirms that density perturbations in the
radiation dominated universe form a field of acoustic waves. In the
flat universe the density perturbations of all length-scales move with
the same sound speed $v=1/\sqrt{3}$. Short and long perturbations do not
form different classes of solutions. Perturbations' velocity is
independent of the wave number, and in particular is the same for
subhorizon and superhorizon inhomogeneities. Although the
gauge-invariant theory confirms the fundamental properties of acoustic
field, which have been known from gauge-specific descriptions
\cite{sachs&wolfe,lukash,golda&woszczyna}, the
propagation equations are not the same. An outstanding property of the
gauge-invariant description is that propagation equation for sound in
radiational era are identical with the propagation equations for
gravitational or electromagnetic waves in the matter dominated universe.

In the open universe perturbations evolve in a more complex manner. The
negative space curvature causes the dispersion of acoustic waves. The
universe geometry determines the minimal frequency for traveling
acoustic waves in the similar way as the geometry of the wave conductor
determines the minimal frequency for waves propagating inside. The
critical frequency is related solely to the space curvature (not to the
Jeans length-scale). Below this frequency perturbations form standing
waves of supercurvature scale. In the radiation dominated universe the
distinction between travelling and standing acoustic waves strictly
coincides with the division into subcurvature and supercurvature
inhomogeneities. Supercurvature standing waves are generic solutions.

As commonly expected, the spectrum transfer function depends on the
universe geometry, but differences are essential only in the large scale
limit. In the subcurvature regime {\it generic Gaussian acoustic field\/}
evolve like the acoustic field in the flat universe. Significant
curvature effects may appear in supercurvature scales --- lowest
multipoles in the MBR temperature map.

\section*{Acknowledgements}

We would like to thank Prof. Andrzej Staruszkiewicz for helpful
remarks concerning the harmonic analysis in the Lobachevski
space. This work was partially supported by State Committee for
Scientific Research, project  \grant.

\end{document}